\documentclass[12pt]{article}
\usepackage[square,authoryear]{natbib}
\usepackage{marsden_article}
\usepackage{epstopdf}

\newcommand{\ns}{\mspace{-1.5mu}}             
\newcommand{\ps}{\mspace{1.5mu}}              

\begin{document}

\title
{\Large \bf Parametrization and Stress-Energy-Momentum Tensors   in Metric Field Theories
}

\author{%
{\bf Marco Castrill\'on L\'opez} \\
Departamento de Geometr\' ia y Topolog\' ia\\[-2pt]
Facultad de Ciencias Matem\' aticas \\[-2pt]
Universidad Complutense de Madrid \\[-2pt]
28040 Madrid, Spain
\\
\and 
{\bf Mark J. Gotay} \\
Department of Mathematics\\[-2pt]
University of Hawai`i\\[-2pt]
Honolulu, Hawai`i 96822, USA \\
\and
{\bf Jerrold~E.~Marsden} \\
Control and Dynamical Systems 107-81\\[-2pt] 
California Institute of Technology\\[-2pt]
Pasadena, California 91125, USA\\[12pt]
}
\date{
December 11, 2007}

\thispagestyle{empty}

\maketitle

\begin{abstract}{\footnotesize
We give an exposition of the parametrization method of \cite{Kuchar1973} in the context of the multisymplectic approach to field theory, as presented in \cite{GoMa2008a}. The  purpose of the formalism developed herein is to make any classical field theory, containing a metric as a sole background field, generally covariant (that is, \emph{parametrized}, with the spacetime diffeomorphism group as a symmetry group) as well as fully dynamic. This is accomplished by introducing certain ``covariance fields''  as genuine dynamic fields.  As we shall see, the multimomenta conjugate to these new fields form the Piola--Kirchhoff version of the stress-energy-momentum tensor field,  and their Euler--Lagrange equations are vacuously satisfied. Thus, these fields have no additional physical content; they serve only to provide an efficient means of parametrizing the theory.
Our results are illustrated with two examples, namely an electromagnetic field and a Klein--Gordon vector field, both on a background spacetime.}
\end{abstract}

\newpage

\section{Introduction.} When one is dealing with classical field theories on a spacetime, the metric may appear as a given background field or it may be a genuine dynamic field satisfying the Einstein equations. The latter theories are often generally covariant, with the spacetime diffeomorphism group as symmetry group, but the former often are considered to have only the isometry group of the metric as a symmetry group.  However, \cite{Kuchar1973} (see also \cite{IsKu1985}) indicated how theories with a background metric can be {\bfi parametrized}, that is, considered as theories that are fully covariant, if one introduces the diffeomorphisms themselves as dynamic fields. The goal of this paper is to develop this idea in the context of multisymplectic classical field theory and to make connections with stress-energy-momentum (``SEM'') tensors. As we shall see, the multimomenta conjugate to these new {\bfi covariance fields} form, to borrow a phrase from elasticity theory, the Piola--Kirchhoff version of the SEM tensor,  and their Euler--Lagrange equations are vacuously satisfied by virtue of the fact that the SEM tensor is covariantly conserved. Thus these fields have no physical content; they serve only to provide an efficient way of parametrizing a field theory. Nonetheless, the resulting generally covariant field theory has several attractive features, chief among which is that it is fully dynamic---all fields satisfy Euler--Lagrange equations. Structurally, such theories are much simpler to analyze than ones with absolute objects or noncovariant elements.

We emphasize that the results of this paper are for those field theories whose Lagrangians are built from dynamic matter or other fields and a non-dynamic background metric. One of our motivations was to find a way to treat background fields and dynamic fields in a unified way in the context of the adjoint formalism. Many of the ideas are applicable to a wider range of field theories, as \cite{Kuchar1973} already indicates, but in this paper we confine ourselves to this important class. The general case is presented in \cite{GoMa2008b} along with a more  detailed discussion of parametrization theory and related topics.

\section{The Covariance Construction.} 

Suppose that we have a metric field theory in which the metric is an absolute object in the sense of \cite{Anderson1967}. For instance, one might consider a dynamic electromagnetic field propagating on a Schwarzschild spacetime. Such a theory is not generally covariant, because the spacetime is fixed, and not all fields are on an equal footing, as the electromagnetic field is dynamic while the gravitational field is not. A somewhat different example is provided by Nordstr$\o$m's theory of gravity (see \S17.6 of \cite{MiThWh1973}), which is set against  a Minkowskian background.

In this section we  explain how to take such a system and construct from it an equivalent field theory that achieves the following goals:
\begin{description}
\item[\hspace{.75in} (I)] The new field theory is generally covariant, and 
\item[\hspace{.75in} (II)] All fields in the new field theory are dynamic. 
\end{description}
This ``covariance construction'' is an extension and refinement of the param\-etrization procedure introduced by  \cite{Kuchar1973}.

\paragraph{\large Setup.} As usual for a first order classical field theory, we start with a bundle $Y \rightarrow X $ whose sections, denoted $\phi$, are the fields under consideration. The dimension of $X$ is taken to be $n + 1$, and we suppose that $X$ is oriented. Let 
\[
\mathcal{L} : J^1 Y \to \Lambda^{n+1} X
\]
be a Lagrangian density for this field theory, where $J^1 Y$ is the first jet bundle of $Y $ and $\Lambda^{n+1} X$ is the space of top forms on $X$. Loosely following the notation of \cite{GoMa1992} or \cite{GoMa2008a}, we write coordinates for $ J^1 Y$ as $ \left( x^\mu, y^A, y^A{}_{\mu} \right)$. In addition, in coordinates,  we shall write 
$$\mathcal{L} = L\! \left( x^\mu, y^A, y^A{}_{ \mu} \right) \! d^{\ps n+1} \ns x.$$
Evaluated on the first jet prolongation of a section $\phi$, the Lagrangian becomes a function of $  \left( x^\mu, \phi^A, \phi^A{}_{, \mu}\right)$; we shall abbreviate this when convenient and simply write $\mathcal{L} ( j^1\ns\phi )$. We assume that the fields $\phi$ are dynamic.

\paragraph{Example.} We will intersperse the example of electromagnetism throughout the paper to illustrate our results. Then $Y$ is the cotangent bundle of 4-dimensional spacetime $X$, sections of which are electromagnetic potentials $A$. The corresponding Lagrangian is written below. \hfill $\blacklozenge$

\paragraph{\large A First Attempt at General Covariance.} Suppose that the spacetime $X$ comes equipped with a fixed, background metric $g$.  The obvious first step in attaining general covariance is to allow $g$ to vary; thus the metric will now be regarded as a genuine \emph{field} $G$ on $X$. (When the metric is regarded as variable, we denote it by $G$, and when we want to revert to its fixed value we use $g$.) So we are led to view the Lagrangian density as a map
\[
\mathcal{L} : J^1 Y \times {\rm Lor}(X) \to \Lambda^{n+1} X
\]
where $ {\rm Lor}(X)$ is the bundle whose sections are  Lorentz metrics on $X$. We correspondingly write $\mathcal{L} ( j^1\ns \phi\, ; G )$; the semicolon is used to separate the dynamic from the nondynamic fields. (We emphasize that $G$ being variable does not mean that it is dynamic; we discuss this point momentarily.) Notice that we have tacitly assumed that the dependence of $\mathcal{L}$ on the metric is pointwise---that is, we have non-derivative coupling. (The more general case of derivative coupling will be considered in \S5. In any event, we remark that derivatively-coupled theories are  considered by many to be pathological.) 

\paragraph{Example.}  The electromagnetic Lagrangian density 
 \[ \mathcal{L}\colon J^1(T^*X) \times \mathrm{Lor}(X)\to \Lambda ^{4} X\]
  is
 \begin{equation}
\mathcal{L}(j^1 \ns A \, ; G )= -\frac14 G^{\mu\alpha}G^{\nu\beta}F_{\alpha\beta} F_{\mu \nu}\sqrt{-G} \, d^{\ps 4}\ns x
\label{oldEMLag}
\end{equation}
where $F_{\mu \nu} = A_{\nu, \mu} - A _{\mu, \nu}.$
\hfill $\blacklozenge$
\bigskip 

Next, assume that the given Lagrangian density $\mathcal{L}$ has the following (eminently reasonable) covariance property for a diffeomorphism $\sigma: X \rightarrow X $:
\begin{equation}
\label{cov}
\sigma _{\ast}\ns  \left( \mathcal{L} (  j^1\ns\phi\, ; G ) \right) = \mathcal{L} \big( j^1\ns(\sigma_Y( \phi))\, ; \sigma_{\ast} G \big)
\end{equation}
where we assume that a way to lift the spacetime diffeomorphism $\sigma$  to a bundle automorphism $\sigma_Y$ of $Y$ has been chosen.

\paragraph{Example.}  For  the electromagnetic 1-form potential $A$, we take the lift to be push-forward on the fiber, which makes it obvious that \eqref{cov} holds in this case.
\hfill $\blacklozenge$

\bigskip

When condition \eqref{cov} holds, we say that the theory is {\bfi generally covariant}, i.e., the Lagrangian density is ${\rm Diff}(X)$-equivariant. Thus we have accomplished objective (I). However, the reader may well remark that this was `too easy,' and would be quite right. The problem is  that it is not clear how, or even {\it if}, $G$ can now be made dynamic. Certainly, $G$ cannot be taken to be variational unless one adds a source term to the Lagrangian density for $G$, for otherwise
\[
\frac{\partial L}{\partial G_{\mu\nu}} = \frac{\delta L}{\delta G_{\mu\nu}} = 0
\] 
as the metric non-derivatively couples to the other fields.
But what should this source term be? If $G$ is gravity, we could use the Hilbert Lagrangian, but otherwise this is unclear.

\paragraph{\large The Covariance Field.} 
The solution to our problem requires more subtlety. We will sidestep both the issues of making $g$ variable, and then making $G$ dynamic, in one fell swoop as follows. We introduce an entirely new field, the ``covariance field'' into the theory. It will `soak up' the arbitrariness in $G$, and will be dynamic. In this way we are able to generate a new generally covariant field theory, physically equivalent to the original one, in which all fields are dynamic. Here is the construction.  

The key idea is to introduce a copy $(S,g)$ of spacetime 
into the fiber of the configuration bundle. Consider (oriented) diffeomorphisms $\eta: X \to S$, thought of as sections of the bundle $S \times X \to X$. We regard the diffeomorphisms $\eta$ as new fields, and correspondingly replace the configuration bundle by $\widetilde Y = Y \times_X (S \times X) \to X$.
Next, modify $\mathcal{L}$ to get the new Lagrangian $\widetilde{\mathcal{L}}$ defined on $J^1\widetilde Y$:
\begin{equation} \label{newLag}
\widetilde{\mathcal{L}} (  j^1\ns\phi,  j^1\ns\eta ) = \mathcal{L} (  j^1\ns\phi; \eta ^{\ast} g ).
\end{equation}
Thus, we obtain a modified field theory with the underlying bundle $\widetilde Y$. The general set up is shown in the figure below. 

\begin{figure}[ht]
\begin{center}
\includegraphics[scale=0.9,angle=0]{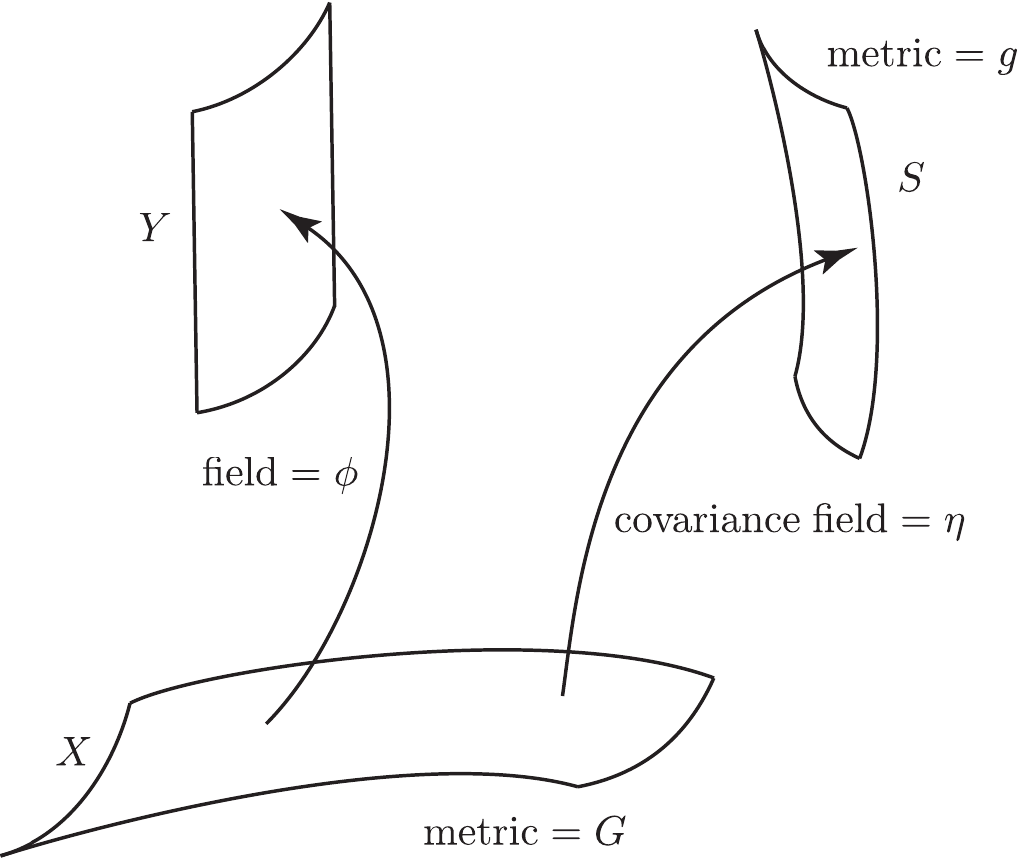}
\end{center}
\end{figure}
\vskip -40pt

\begin{center}
{\footnotesize The general set up for the introduction of covariance fields.}
\end{center}
\bigskip

 Let coordinates on $S$ be denoted $u^a$ and the associated jet coordinates be denoted $u ^a {} _{\mu}$. Then, writing $\mathcal L = L\, d^{\ps 4}x$ and similarly for $\widetilde {\mathcal L}$,  in coordinates equation \eqref{newLag} reads 
 \begin{equation}\label{widetL}
\widetilde{{L}}
\left( x^\mu, y^A, y^A{}_{\mu}, u^a, u ^a {} _{\mu} \right)
= L\! \left( x^\mu, y^A, y^A{}_{\mu}\, ; G _{\mu \nu} \right)\! ,
\end{equation}
where  from the definition of pull-back
\[
( \eta ^{\ast} g )_{\mu\nu } (x ) =  \eta ^{a}{}_{, \mu} (x) \eta ^{b}{}_{, \nu} (x) g _{a b } ( \eta (x) )
\]
we obtain
 \begin{equation}\label{G}
G_{\mu\nu} = u ^a {} _{\mu} u ^b {} _{\nu} \, g _{a b }.
\end{equation}
From \eqref{widetL} one verifies that the Euler--Lagrange equations for the fields $\phi^A$ remain unchanged. 

\paragraph{Example.} For the electromagnetic field, our construction produces
 \begin{equation}
\tilde{\mathcal L}(j^1\ns A ,j^1 \eta )= -\frac14 g^{ac} g^{bd} \kappa^\mu{}_c \kappa^\alpha{}_a \kappa^\nu{}_d \kappa^\beta{}_b\ps  F_{\mu\nu} F_{\alpha\beta} \, \sqrt{-g}\, (\det \eta_*) \, d^{\ps 4}\ns x
 \label{newEM Lag}
\end{equation}
where $\eta_*$ is the Jacobian of $\eta$ and $\kappa = \eta^{-1}.$
\hfill $\blacklozenge$
\bigskip

We pause to point out the salient features of our construction. First, the fixed metric $g$ on spacetime is no longer regarded as living on $X$, but rather on the copy $S$ of $X$ in the fiber of the configuration bundle $\widetilde Y$. So $g$ is no longer considered to be a field---it has been demoted to a mere geometric object on the fiber $S$. Second, the variable metric $G$ on $X$ is identified with $\eta^*g$, and thus acquires its variability from that of $\eta$. So $G$ as well is no longer a field per se, but simply an abbreviation for the quantity $\eta^*g$. 
Finally, we gain a field $\eta$ which we allow to be dynamic; in the next subsection we will see that this imposes no restrictions on the theory at all.

The first  key observation is that the modified theory is indeed generally covariant. To this end, recall that, as was explained earlier, given  $\sigma \in \mathrm{Diff}X$, there is assumed to be a lift $\sigma _Y \colon Y \to Y$. For the trivial bundle $S \times X$, we define
\begin{eqnarray}
\sigma _{S}\colon S \times X &\to& S \times X \nonumber \\
(u,x)& \mapsto & (u,\sigma (x)).
\label{psiS}
\end{eqnarray}

\begin{theorem}
The Lagrangian density $\widetilde{\mathcal{L}}\colon J^1(Y \times_X (S\times X)) \to \Lambda ^{n+1}X$ is $\operatorname{Diff} (X) $-equivariant,  that is,
\[
\sigma _{\ast} \Big( \widetilde{\mathcal{L}} \big( j^1\ns\phi,  j^1\ns\eta \big) \Big)
= \widetilde{\mathcal{L}} \left( j^1\ns( \sigma_Y(\phi)), j^1\ns( \sigma_S(\eta) )\right).
\]
\end{theorem}

\begin{proof}
This is an easy consequence of  the definitions \eqref{newLag} and \eqref{psiS},  and the covariance assumption  \eqref{cov}. Indeed
\begin{align*}
\widetilde{\mathcal{L}} \left( j^1\ns( \sigma_Y(\phi)),  j^1\ns( \sigma_S(\eta) )\right) 
&= 
{\mathcal{L}} \left( j^1\ns( \sigma_Y(\phi))\, ; (\eta \circ \sigma^{-1} )^*g\right) \\[1.5ex]
&= {\mathcal{L}} \left( j^1\ns( \sigma_Y(\phi))\, ; (\sigma^{-1} )^*(\eta^* g)\right) \\[1.5ex]
&=  \sigma_*\!\left(\mathcal{L}(j^1 \ns \phi)\, ; (\eta ^* g))\right)\\[1.5ex]
&= \sigma_*\!\left(\widetilde {\mathcal{L}} (j^1 \ns \phi, j^1 \ns\eta )\right).
\end{align*} \vskip -24pt
\end{proof} 
Because of this property, we call $\eta$ the {\bfi covariance field}.

\paragraph{Example.} From \eqref{newEM Lag}
 it is clear that the modified electromagnetic theory is generally covariant.
 \hfill $\blacklozenge$

\section{The Dynamics of the Covariance Field.}
Next we will show something remarkable: the Euler--Lagrange equation for the covariance field $\eta$ is vacuous. This is the main reason that, in the present context, we can introduce $\eta$ as a dynamic field with impunity, namely, its Euler--Lagrange equation does not add any new information to, or impose any restrictions upon, the system.  Since, as we mentioned earlier, the Euler--Lagrange equations for the fields $\phi^A$ remain unaltered, we see that \emph{the parametrized system is physically equivalent to the original system.}

First we compute the multimomenta conjugate to the field $\eta$ for the parame\-trized field theory with Lagrangian $\widetilde{\mathcal{L}}$. Recall that in multisymplectic field theory, the multimomenta conjugate to the multivelocities $u^A{}_{\mu}$ 
are defined by 
\begin{equation*}
\rho _a {} ^{\mu} =  \frac{\partial \widetilde{L } }{\partial u ^a {} _{\mu}}.
\end{equation*}
Using the chain rule together with the relations \eqref{widetL} and \eqref{G}, we find that 
\begin{equation} \label{pkmom}
\rho _a {} ^{\mu} =  2 \ps \frac{\partial L }{\partial G _{\mu \nu } } \ps u ^b{} _{\nu } g _{ab}.
\end{equation}
Recall from \cite{GoMa1992} that, as we have assumed that $G$ is the only nondynamic field, and does not derivatively couple to the other fields, the SEM tensor density for the \emph{original} system with Lagrangian $L$ and metric $G$ is given by the Hilbert formula:
\begin{equation} \label{hsemdef}
\mathfrak T ^{\mu \nu} =  2\ps \frac{\delta L }{\delta G _{\mu \nu } } = 2\ps \frac{\partial L }{\partial G _{\mu \nu } }.
\end{equation}
From \eqref{pkmom} we conclude that the multimomenta conjugate to to the covariance field $\eta$ are given by
the {\bfi Piola-Kirchhoff SEM tensor density}:
\begin{equation*}
\rho _a {} ^{\mu} =  \mathfrak T ^{\mu \nu} u ^b{} _{ \nu } g _{ab}.
\end{equation*}
This is a familiar object in elasticity theory. Observe that $\rho_a{}^\mu$ is a two-point tensor density: it has one leg ($a$) in the spacetime $S$ in the fiber---analogous to the spatial representation in elasticity theory, and the other leg ($\mu$) in the spacetime $X$ in the base---analogous to the material representation.

Now we compute the Euler--Lagrange equations for the $\eta^a$. These are:
\[
\frac{\partial \tilde{L} }{\partial \eta ^a} - \frac{\partial}{\partial x^\mu}
\left( \frac{\partial \tilde{L} }{\partial \eta ^a{} _{,\mu}}\right)=0
\]
for $a=1,\ldots,\operatorname{dim}X$. Expanding the derivatives via the chain rule and using the same type of calculation as in the derivation of \eqref{pkmom} to write the equations in terms of $L$ rather than $\tilde{L} $,  the preceding equation becomes
\[
\frac{\partial L}{\partial G_{\mu \nu}}\ps \eta^c{}_{,\mu}\eta^d{}_{,\nu}\ps \frac{\partial g_{cd}}
{\partial u^a}- 2\ps \frac{\partial}{\partial x^\mu}\ns \left(\frac{\partial L}
{\partial G_{\mu \nu}}\ps \eta ^c{}_{,\nu}g_{ac}\right)=0.
\]
Replacing $\partial L /\partial G_{\mu \nu}$ by (half of) $\mathfrak T^{\mu \nu}$, and differentiating using the product rule, we obtain
\[
\mathfrak T^{\mu \nu}\eta^c{}_{,\mu} \eta^d{}_{,\nu} \ps \frac{\partial g_{cd}}{\partial u^a}
-2\left(\frac{\partial \mathfrak T^{\mu \nu}}{\partial x^\mu}\ps \eta^c{}_{,\nu}g_{ac}+\mathfrak T^{\mu \nu}
\eta^c{}_{,\mu \nu}g_{ac}+\mathfrak T^{\mu \nu}\eta ^c{}_{,\nu}\ps \frac{\partial g_{ac}}{\partial u^d} \ps \eta^d{}_{,\mu}\right)=0,
\]
for $a=1,\ldots, \dim X$.

Multiplying by the inverse matrix $g^{ab}$ one gets
\[
\mathfrak T^{\mu \nu}\eta^c{}_{,\mu} \eta^d{}_{,\nu} \ps \frac{\partial g_{cd}}{\partial u^a}\ps g^{ab}
-2\left(\frac{\partial \mathfrak T^{\mu \nu}}{\partial x^\mu}\ps \eta ^b{}_{,\nu}+\mathfrak T^{\mu \nu}
\eta^b{}_{,\mu \nu}+\mathfrak T^{\mu \nu}\eta^c{}_{,\nu}\eta^d{}_{,\mu}\frac{\partial g_{ac}}{\partial u^d}\ps g^{ab}\right)=0,
\]
for $b=1,\ldots,\dim X$. And now, we multiply by $\kappa^\rho{}_b$, the inverse matrix of the Jacobian
$\eta ^b{}_{,\nu}$
\[
\mathfrak T^{\mu \nu}\eta^c{}_{,\mu} \eta^d{}_{,\nu} \ps \frac{\partial g_{cd}}{\partial u^a}\ps g^{ab}\kappa^\rho{}_b
-2\left(\frac{\partial \mathfrak T^{\mu \rho}}{\partial x^\mu}+\mathfrak T^{\mu \nu}\eta ^b{}_{,\mu \nu} \kappa^\rho{}_b
+\mathfrak T^{\mu \nu}\eta^c{}_{,\nu}\eta^d{}_{,\mu}\ps \frac{\partial g_{ac}}{\partial u^d}\ps
g^{ab}\kappa^\rho{}_b \right)=0,
\]
for $\nu =1,\ldots,\dim X$. Taking into account the symmetry $\mathfrak T^{\mu \nu}=\mathfrak T^{\nu \mu}$, the preceding equation becomes
\begin{align*}
\mathfrak T^{\mu \nu}\eta^c{}_{,\mu} \eta^d{}_{,\nu} \kappa^\rho{}_b
\left( \frac{\partial g_{cd}}{\partial u^a}\ps g^{ab} 
\right. & \left. - \ \frac{\partial g_{ad}}
{\partial u^c}\ps g^{ab} - \frac{\partial g_{ac}}{\partial u^d}\ps g^{ab} \right) \\[2ex]
& - \ 2 \left(\frac{\partial \mathfrak T^{\mu \rho}}{\partial x^\mu}+\mathfrak T^{\mu \nu}\eta ^b{}_{,\mu \nu} \kappa^\rho{}_b \right) = 0.
\end{align*}
Recalling the expression of the Christoffel symbols of the metric $g$, namely,
\[
\gamma ^b _{cd}=\frac{1}{2}g^{ab}\left(\frac{\partial g_{ac}}{\partial u^d}+\frac{\partial g_{ad}}{\partial
u^c}-\frac{\partial g_{cd}}{\partial u^a}\right),
\]
we obtain
\begin{equation} \label{etaeqn}
- 2 \mathfrak T^{\mu \nu}\eta^c{}_{,\mu} \eta^d{}_{,\nu} \gamma ^b _{cd}\ps \kappa^\rho{}_b
-2\left(\frac{\partial \mathfrak T^{\mu \rho}}{\partial x^\mu}+\mathfrak T^{\mu \nu}\eta ^b{}_{,\mu \nu} \kappa^\rho{}_b \right)  =0.
\end{equation}
Finally, recall how the Christoffel symbols $\gamma ^b _{cd}$ for $g$ and the symbols $\Gamma ^\rho _{\mu \nu}$ for $G=\eta ^* g$ are related:
\begin{equation} 
\Gamma ^\rho _{\mu \nu} = \frac{\partial ^2u^b}{\partial x^\mu \partial x^\nu}\frac{\partial x^\rho} {\partial u^b} + \frac{\partial u^c}{\partial x^\mu}\frac{\partial u^d}{\partial x^\nu}
\ps \gamma ^b _{cd} \ps  \frac{\partial x^\rho}{\partial u^b}.
\label{Gg}
\end{equation}
Using this in \eqref{etaeqn} gives
\[
-2\left(\frac{\partial \mathfrak T^{\mu \rho}}{\partial x^\mu} + \mathfrak T^{\mu \nu}\Gamma ^\rho _{\mu \nu}\right)=0,
\]
for $\nu =1,\ldots,\dim X$, which is exactly the vanishing of the  covariant divergence of the tensor \emph{density} $\mathfrak T^{\mu \nu}$.
 
Thus, we have proven the following basic result.

\begin{theorem} The Euler--Lagrange equations for the covariance field $\eta$ are that the covariant divergence of the SEM tensor density $\mathfrak T ^{\mu \nu}$ is zero.
\end{theorem}

It is known from Proposition 5 in \cite{GoMa1992} that the SEM tensor is covariantly conserved when the metric $G$ is the \emph{only} nondynamic field. Thus, in our context, the equation $\nabla_\mu \mathfrak T^{\mu\nu} = 0$ is an identity, whence

\begin{corollary} The Euler--Lagrange equations for the covariance field $\eta$ are vacuously satisfied.
\end{corollary}

Consequently the covariance field has no physical import. We are free to suppose $\eta$ is dynamic, and so we have accomplished goal (II): we have constructed a new field theory in which \emph{all} fields are dynamic. 

\section{The SEM Tensor.}

It is interesting to compare the SEM tensors for the original and parame\-trized systems. In \cite{GoMa1992} the SEM tensor density $\mathfrak T^\mu{}_\nu$ is defined in terms of fluxes of the multimomentum map $J^{\mathcal L}$ associated to the action of the spacetime diffeomorphism group. We rapidly recount some of the basic ideas.

Consider the lift of an infinitesimal diffeomorphism $\xi \in \mathfrak{X}(X)$ to $Y$; it can be expressed
\[
\xi _Y = \xi ^{\mu} \frac{\partial}{\partial x^{\mu}} + \xi ^A \frac{\partial}{\partial y ^A}
\]
where we suppose that
\begin{equation*}
\xi ^A = C^{A\rho _1 \dots \rho _k}_{\ \ \nu} \xi ^{\nu}{}_{,\rho _1 \ldots \rho _k} + \ldots +
C^{A\rho}_{\ \ \nu} \xi^{\nu}{}_{,\rho}+C^A{}_{\nu} \xi ^{\nu}
\end{equation*}
for some coefficients $C^{A\rho _1 \dots \rho _k}_{\ \ \nu} ,  \ldots ,C^{A\rho}_{\ \ \nu} , C^A{}_{\nu}$. The largest value of $k$ for which one of the top coefficients $ C^{A\rho _1 \dots \rho _k}_{\ \ \nu} $ is nonzero is the {\bfi differential index} of the field theory. We assume henceforth that  the index $\leq 1$---the most common and important case (e.g., when the fields are all tensor fields).

In this context, Theorem 1 along with Remark 4 of \cite{GoMa1992} shows that the SEM tensor density $\mathfrak T$ for a Lagrangian density ${\mathcal{L}}$ is uniquely determined  by
\begin{equation}\label{DefSEM} 
\int_ \Sigma i ^\ast _\Sigma (j ^1\ns \phi) ^\ast J ^{ \mathcal{L}} (\xi _Y) = \int _\Sigma
\mathfrak{T} ^\mu {}_{\nu} (\phi) \xi ^\nu d ^{\ps n} \ns x _\mu 
\end{equation} 
for all vector fields $ \xi$ on $X$  with compact support and all hypersurfaces $ \Sigma $, where $ i _
\Sigma : \Sigma \rightarrow X $ is the inclusion. 
The multimomentum map $J ^{ \mathcal{L}} $ gives, roughly speaking, the flow of momentum and energy through spacetime; according to the quoted theorem, the  fluxes of this flow across hypersurfaces are realized via the SEM tensor density.

Manipulation of \eqref{DefSEM} (see formula (3.12) of \cite{GoMa1992}) shows that $\mathfrak T$  is given by
\[
{\mathfrak{T}}^{\mu}{}_{\nu} ={L}\delta ^{\mu}{}_{\nu} - \frac{\partial{L}}{\partial \psi ^A{}_{,\mu}} \ps
\psi ^A{}_{,\nu} + \frac{\partial{L}}{\partial \psi ^A{}_{,\mu}} \ps C^A{} _\nu +
D_{\rho} \!\left(\frac{\partial{L}}{\partial \psi ^A{}_{,\rho}}\ps C^{A\mu}_{\ \ \nu}\right) 
\]
where the summation extends over \emph{all} fields $\psi^A$.

We apply this to the newly parametrized theory. Note that if the index of the original theory is $\leq 1$, then that for the parametrized theory will be also. As well
from \eqref{psiS} we see that the lift of $\xi$ to $S \times X$ is trivial:
\[
\xi ^{a} = 0,
\]
that is, there are no terms in the $\partial /\partial u^a$ directions in $\xi_{\widetilde Y}$. Thus the corresponding coefficients $C^{a\cdots}_{\  \ \nu}$ all vanish.
The SEM tensor for $\widetilde{\mathcal{L}}$ therefore reduces to
\[
\widetilde{\mathfrak{T}}^{\mu}{}_{\nu} = \widetilde{L}\delta ^{\mu}{}_{\nu} - \frac{\partial \widetilde{L}}{\partial \phi ^A{}_{,\mu}} \ps
\phi ^A{}_{,\nu} + \frac{\partial \widetilde{L}}{\partial \phi ^A{}_{,\mu}} \ps C^A{} _\nu +
D_{\rho} \! \left(\frac{\partial \widetilde{L}}{\partial \phi ^A{}_{,\rho}}\ps C^{A\mu}_{\ \ \nu}\right) - \frac{\partial \widetilde{L}}
{\partial \eta ^a{}_{,\mu}}\ps \eta ^a{}_{,\nu}.
\]
On the other hand,
\[
\frac{\partial \widetilde{L}}{\partial \eta ^a{}_{,\mu}}\ps \eta ^a{}_{,\nu} = 2 \frac{\partial L}{\partial G_{\mu\rho}} \ps \eta ^b{}_{,\rho}\,g_{ab}\,\eta ^a{}_{,\nu}
= 2 \ps \frac{\partial L}{\partial G_{\mu\rho}} \ps G_{\rho \nu} 
\]
and
\[
\frac{\partial \widetilde{L}}{\partial \phi ^A{}_{,\mu}} = \frac{\partial L}{\partial \phi ^A{}_{,\mu}},
\]
so that we can write
\begin{equation*}
\widetilde{\mathfrak{T}}^{\mu}{}_{\nu} = L\delta ^{\mu}{}_{\nu} - \frac{\partial L}{\partial \phi ^A{}_{,\mu}}\ps
\phi ^A{}_{,\nu} + \frac{\partial L}{\partial \phi ^A{}_{,\mu}} \ps C^A{}_\nu +
D_{\rho}\! \left(\frac{\partial L}{\partial \phi ^A{}_{,\rho}}\ps C^{A\mu}_{\ \ \nu}\right) - 2 \frac{\partial L}{\partial G_{\mu\rho}}\ps  G_{\rho \nu}.
\end{equation*}
But the first four terms on the RHS of this equation comprise the SEM tensor density of the original theory since the $G_{\mu\nu}$ do not derivatively couple to the $\phi^A$ (cf. eqn. (4.4) in \cite{GoMa1992}).
 Thus the SEM tensor densities of the original and parametrized systems are related according to: 
 \begin{proposition}
\begin{equation*}\label{SEM2}
\widetilde{\mathfrak{T}}^{\mu}{}_{\nu} = \mathfrak{T}^{\mu}{}_{\nu} -  2 \frac{\partial L}{\partial G_{\mu\rho}} \ps G_{\rho \nu}.
\end{equation*}
\end{proposition}

But then $\widetilde{\mathfrak{T}}^{\mu}{}_{\nu} = 0$ on shell by the Hilbert formula \eqref{hsemdef}. 
Therefore, we explicitly see that the SEM tensor density for the fully covariant, fully dynamic modified theory vanishes. 
One can also obtain this result directly by applying the generalized Hilbert formula (3.13) in \cite{GoMa1992} to the parametrized theory,  since it is fully dynamic.

\paragraph{Example.} In the case of electromagnetism, one may compute directly from \eqref{newEM Lag} that $\widetilde{\mathfrak{T}}^{\mu}{}_{\nu} = 0$. One could also compute from \eqref{oldEMLag} that 
\begin{equation*}
{\mathfrak{T}}^{\mu}{}_{\nu}  = -\left(\frac14 \delta^{\mu}{}_{\nu} F_{\alpha\beta}F^{\alpha\beta} + F^{\alpha\mu}F_{\nu\alpha}\right)\! \sqrt{-G} = 2 \ps \frac{\partial L}{\partial G_{\mu\rho}} \ps G_{\rho\nu}.
\end{equation*}
\vskip -32pt
 \hfill $\blacklozenge$

\section{Derivative Couplings.}
Here we briefly consider the situation, although perhaps exotic, when the metric  derivatively couples to the other fields. For simplicity, however, we suppose the theory remains first order. So  the Lagrangian density is taken to be  a map
\[
\mathcal{L} : J^1\ns \big(Y \times_X {\rm Lor}(X)\big) \to \Lambda^{n+1} X.
\]

As before, modify $\mathcal{L}$ to get the new Lagrangian $\widetilde{\mathcal{L}}$ defined on $J^2\widetilde Y$:
\begin{equation*} 
\widetilde{\mathcal{L}} \big(  j^1\ns\phi,  j^2 \eta \big) = \mathcal{L} \big(  j^1\ns\phi \ps ; j^1\ns (\eta ^{\ast} g) \big).
\end{equation*}
(Since $\eta^* g$ depends upon the first derivatives of $\eta$, $j^1(\eta^* g)$ will depend upon its second derivatives.
Thus, we obtain a modified \emph{second} order field theory
 with the underlying bundle $\widetilde Y$.) The discussion proceeds as in the above, with only obvious changes. In particular, if $\mathcal L$ is Diff$(X)$-covariant, then so is $\widetilde{\mathcal L}$.

\paragraph {\bf Example.} As a simple illustration of a derivatively coupled theory, consider a vector meson with mass $m$. Then $Y$ is the tangent bundle of spacetime, and its sections $\phi^{\ps \mu}$ are Klein--Gordon vector fields. 
The Lagrangian density is the map
 \[ 
 \mathcal{L}\colon J^1\ns \big(TX  \times_X \mathrm{Lor}(X)\big)\to \Lambda ^{4} X
 \]
defined by
 \begin{equation*}
\mathcal{L}(j^1\ns \phi \, ; j^1\ns G )= \frac12 G_{\sigma\ns \rho}\Big(G^{\mu \nu}\phi^\sigma{}_{;\mu}  \ps \phi^{\ps\rho}{}_{;\nu}  - m^2 \phi^\sigma \phi^{\ps\rho} \Big)\sqrt{-G} \, d^{\ps 4}\ns x.
\end{equation*}
where the semicolon denotes the covariant derivative with respect to $G$.

Our construction produces
 \begin{align*}
\tilde{\mathcal L}(j^1 \phi ,j^2 \eta )= &\ \frac12 \eta^c{}_{,\sigma} \eta^d{}_{,\rho} \ps g_{cd}  \bigg(\! \kappa^\mu{}_a \kappa^\nu{}_b \ps g^{ab}\Big[\phi^ \sigma{}_{,\mu} + \big(\eta^ g{}_{,\mu\tau} +\eta^e{}_{,\mu} \ps \eta^f{}_{,\tau}
\ps \gamma ^ g _{ef} \big)  \kappa^ \sigma{}_g\ps \phi^\tau \Big] \nonumber \\[1.5ex]
& \times 
\Big [\phi^{\ps \rho}{}_{,\nu}  + \big(\eta^h{}_{,\nu\xi} + \eta^p{}_{,\nu} \ps \eta^q{}_{,\xi} \ps \gamma ^h _{pq} \big)  \kappa^\rho{}_h\ps \phi^\xi\Big ] \nonumber
\\[1.5ex] 
& - m^2 \phi^\sigma \phi^{\ps \rho} \bigg)\sqrt{-g}\, (\det \eta_*) \, d^{\ps 4}\ns x
\end{align*}
where $\eta_*$ is the Jacobian of $\eta$ and we have made use of \eqref{Gg}. 
\hfill $\blacklozenge$

\bigskip

Now we turn to the Euler--Lagrange equations for the $\eta^a$ which, since $\widetilde{\mathcal L}$ is second order in the $\eta^a$, are:
\[
\frac{\partial \widetilde{L} }{\partial \eta ^a} - \frac{\partial}{\partial x^\mu}
\left( \frac{\partial \widetilde{L} }{\partial \eta ^a{} _{,\mu}}\right) +
 \frac{\partial^2}{\partial x^\nu \partial x^\mu}
\left( \frac{\partial \widetilde{L} }{\partial \eta ^a{} _{,\mu\nu}}\right)
=0
\]
for $a=1,\ldots,\operatorname{dim}X$. The calculation of the LHS is similar to the previous one, but slightly more complicated. 
In any event, we find that $\eta$ satisfies the Euler--Lagrange equations $\iff$ 
$\nabla_\mu \mathfrak T^{\mu\nu} = 0$, where now by the Hilbert formula
$$\mathfrak T^{\mu\nu} =  2\ps \frac{\delta L }{\delta G _{\mu \nu } } = 2\ps \bigg[\frac{\partial L }{\partial G _{\mu \nu }} - \frac{\partial}{\partial x^\rho}  \left(\frac{\partial L }{\partial G _{\mu \nu,\rho } }\right) \bigg].
$$
Thus for 
(first order) derivative couplings the covariance field remains vacuously dynamic. It is likely this will remain true for derivative couplings of arbitrary order, but we have not verified this as yet.


\paragraph{\Large Acknowledgments.} We dedicate this paper to Darryl Holm on
his 60$^{\rm th}$ birthday. We thank him for his interest in the ideas in this paper and
for his many inspiring works over the years. MJG and JEM thank the National Science Foundation for its occasional support of work of this sort. MCL was partially supported by DGSIC (Spain) under grant MTM2007-60017.

\end{document}